\documentclass[%
reprint,
footinbib,
amsmath,amssymb,
pre,
floatfix,
]{revtex4-2}

\usepackage{graphicx}
\usepackage{dcolumn}
\usepackage{bm,upgreek}

\usepackage[colorlinks,
allcolors=blue,
urlcolor=black,
]{hyperref}
\usepackage{color}
\usepackage[capitalise]{cleveref}
\usepackage{natbib}
\usepackage{float}

\graphicspath{{Figures/}}

\newcommand{\vect}[1]{\mathbf{#1}}

\newcommand{\edot}{\dot{e}}
\newcommand{\sdot}{\dot{s}}
\newcommand{\xidot}{\dot{\xi}}
\newcommand{\yy}{\mathrm{y}}
\newcommand{\yc}{\mathrm{yc}}
\newcommand{\ke}{k_\mathrm{e}}
\newcommand{\ce}{c_\mathrm{e}}
\newcommand{\gE}{g_\mathrm{e}}
\newcommand{\gS}{g_\mathrm{s}}
\newcommand{\fy}{f_\yy}
\newcommand{\zz}{\vect{z}}

\newcommand{\units}[1]{~$\mathrm{#1}$}

\providecommand{\Eqref}[1]{\cref{#1}}
\providecommand{\Figref}[1]{Fig.~\ref{#1}}
\providecommand{\Secref}[1]{\Cref{#1}}
\providecommand{\secref}[1]{Sec.~\ref{#1}}
\providecommand{\Appref}[1]{Appendix~\ref{#1}}
\Crefformat{Eqs}{Eqs.~(#2#1#3)}	

\newcommand{\comsol}{\textit{Comsol Multiphysics}}
\newcommand{\ie}{\textit{i.e.}}
\newcommand{\eg}{\textit{e.g.}}
\newcommand{\etc}{\textit{etc}}

\newenvironment{subalign}[1][foolabel]
	{\subequations \label[Eqs]{#1}
	\align}
	{\endalign
	\endsubequations}

\begin{document}
	
\preprint{APS/123-QED}

\title{
	Strongly Nonlinear Wave Propagation in Elasto-plastic Metamaterials: Low-order Dynamic Modeling
	}

\author{Samuel P. Wallen}
\author{Michael R. Haberman}%
\affiliation{%
	Applied Research Laboratories\\
	Walker Department of Mechanical Engineering\\
	The University of Texas at Austin
}%

\author{Washington DeLima}
\affiliation{
	Honeywell FM\&T, KCNSC
}%

\date{\today}

\begin{abstract}
	Nonlinear elastic metamaterials are known to support a variety of dynamic phenomena that enhance our capacity to manipulate elastic waves. Since these properties stem from complex, subwavelength geometry, full-scale dynamic simulations are often prohibitively expensive at scales of interest. Prior studies have therefore utilized low-order effective medium models, such as discrete mass-spring lattices, to capture essential properties in the long-wavelength limit. While models of this type have been successfully implemented for a wide variety of nonlinear elastic systems, they have predominantly considered dynamics depending only on the instantaneous kinematics of the lattice, neglecting history-dependent effects, such as wear and plasticity.	To address this limitation, the present study develops a lattice-based modeling framework for nonlinear elastic metamaterials undergoing plastic deformation. Due to the history- and rate-dependent nature of plasticity, the framework generally yields a system of differential-algebraic equations whose computational cost is significantly greater than an elastic system of comparable size. We demonstrate the method using several models inspired by classical lattice dynamics and continuum plasticity theory, and explore means to obtain empirical plasticity models for general geometries.
\end{abstract}

\maketitle

\section{Introduction}\label{sec:Intro}

	Acoustic and elastic metamaterials are engineered fluids and solids known to manipulate mechanical waves in a myriad of ways that are difficult or unattainable with naturally-occurring materials and conventional composites \cite{Hussein2014review,Haberman2016acoustic,Cummer2016}. While most prior studies have examined linear dynamic behavior, nonlinear metamaterials have also been investigated extensively \cite{Patil2022, jiao2023mechanical} and have been shown to support a variety of rich dynamic phenomena, such as solitary waves and breathers \cite{Boechler2010,Wallen2017breathers,Chong2018,deng2021nonlinear}, buckling instabilities \cite{Nadkarni2014,shan2015multistable, raney2016stable, nadkarni2016unidirectional, deng2017elastic, Kochmann2017rev,jin2020guided, fancher2023,ramakrishnan2023architected,jiao2024phase}, mode conversion \cite{Ganesh2015,Wallen2017}, and nonreciprocity \cite{Liang2010,Boechler2011a,Bunyan2018,Nassar2020review,patil2022leveraging}. Of particular relevance to the present work are nonlinear elastic metamaterials (NLEMs) that have been proposed to mitigate shocks and impacts \cite{Correa2015,restrepo2015phase, hewage2016double, findeisen2017characteristics, bertoldi2017harnessing, 
zhang2019energy,tan2019design,chen2020light, chen2021novel, 
chen2022closed, zhang2021architected,zhang2021architected, ma2022energy,guo2023quasi,begley2024design}, which often leverage intricate lattice structures to achieve superior energy dissipation and redistribution.
	
	Since NLEM properties stem from complex subwavelength geometry, direct numerical simulations at full resolution are often intractable at scales of interest. Therefore, prior studies have turned to effective medium models, which capture essential properties in the long-wavelength limit. One class of such models is a discrete-element model (DEM) comprising a periodic arrangement of lumped mass, spring, and damper elements, whose constitutive relationships are computed to match behaviors observed in experiments or fine-scale simulations of representative volume elements. Using methods of analytical mechanics (\eg, Newton's laws or Lagrangian formalism), the equations of motion of the DEM may be formulated as a system of ordinary differential equations (ODEs).
	
	While DEMs have been studied in various areas of physics for quite some time, they have recently gained popularity as models for NLEMs due to their ability to capture a wide variety of complex physical mechanisms, such as Hertzian contact \cite{Daraio2006,Chong2018}, rotational dynamics \cite{Wallen2017,wallen2017thesis,deng2019focusing}, and elastic instabilities \cite{Nadkarni2014,jiao2024phase}, in a computationally efficient framework. Despite the prevalence of DEM in the NLEM literature, the majority of prior work has considered only recoverable deformation and forces that depend only on the instantaneous kinematics of the system, while history-dependent effects, such as wear and plasticity, have been much less explored. A notable exception is in the field of granular crystals, where elasto-plastic extensions of Hertzian contact theory have been implemented in DEM  and compared with experiments \cite{Burgoyne2014,Burgoyne2015}. Plastic wave propagation has also been treated semi-analytically in one-dimensional (1D) continuum phononic crystals \cite{Dorgant2024}. However, these specific models are of limited applicability for broader classes of NLEM because \textit{i}) plasticity models, when available, incorporate mathematical forms that vary significantly from contact mechanics (see, \eg, \cite{nematnasser2004plasticity} for examples); and \textit{ii}) for general NLEM geometries, analytical elasto-plastic models are not typically available. It is therefore of interest to develop a general DEM approach with applicability to a wide variety of elasto-plastic wave propagation problems.
	
	In this work, we develop a 1D DEM framework for general classes of nonlinear elastic materials undergoing plastic deformation and demonstrate its implementation using several elasto-plastic constitutive models, including an empirical model derived from finite element analysis (FEA) of a representative unit cell geometry. The remaining sections are outlined as follows: \Secref{sec:SingleOsc} introduces the elasto-plastic DEM for an isolated spring-mass oscillator, develops its equations of motion, and discusses their numerical solution. \Secref{sec:Chains} generalizes the DEM for periodic chains of elasto-plastic elements and demonstrates its usage via several elasto-plastic constitutive models from existing literature, including a rate-dependent model and a plastic extension of the soliton-bearing Toda lattice \cite{Toda1970}. \Secref{sec:NSH} demonstrates a procedure to generate an elasto-plastic constitutive relationship from cyclic force-displacement data, using a negative-stiffness honeycomb geometry as an example, and implements it in the DEM framework. Finally, \secref{sec:Conclusion} concludes the work.
	
\section{Individual Oscillator}\label{sec:SingleOsc}
	
\begin{figure}[h]
				\centering
		\includegraphics[width=0.7\linewidth]{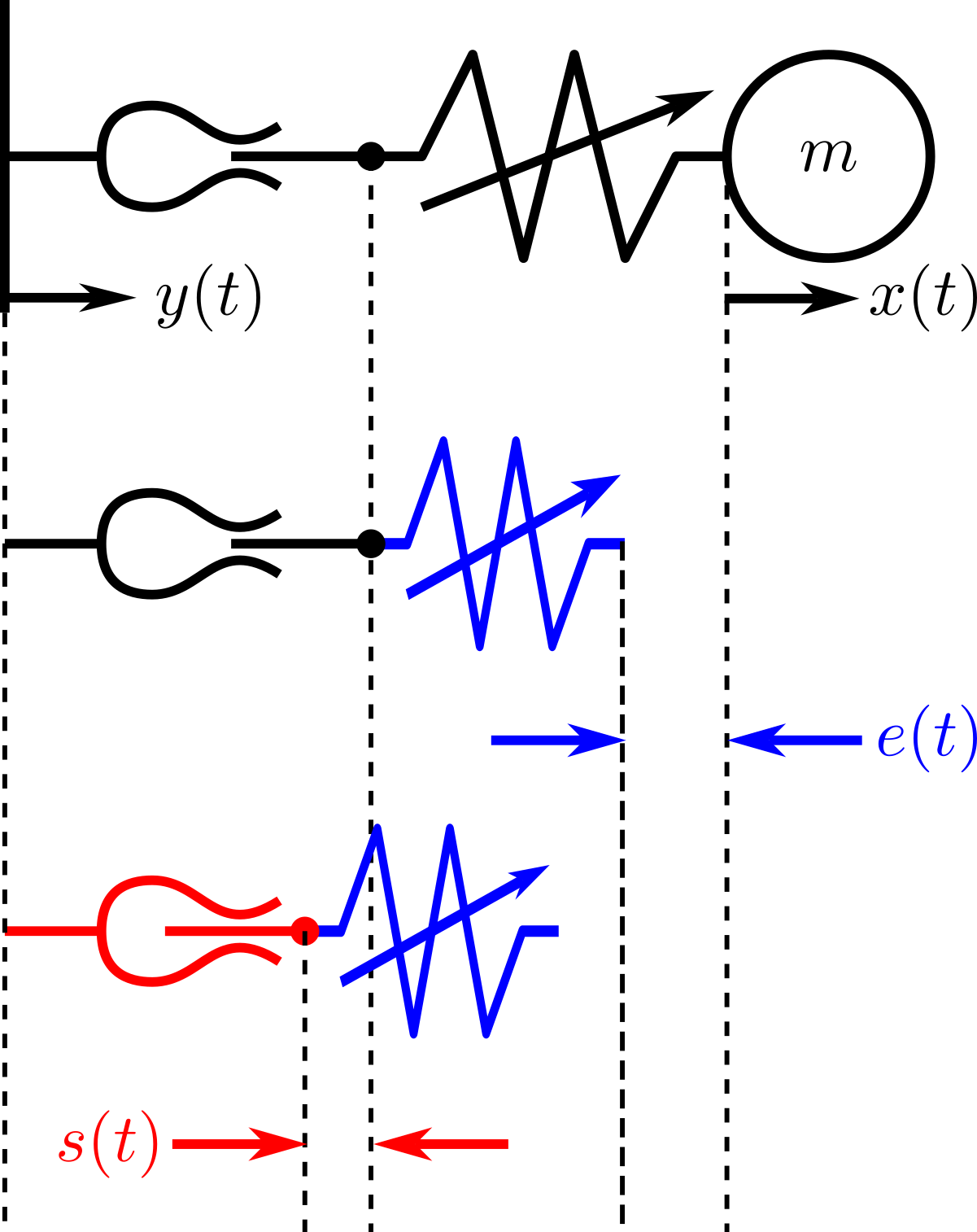}
		\caption{Schematic of a spring-mass oscillator exhibiting nonlinear, elasto-plastic deformation. The blue nonlinear spring deforms elastically in either direction, while the red slider deforms plastically and in compression only. Since the elastic and plastic elements may deform independently, the oscillator possess two kinematic degrees of freedom.}
		\label{fig:SDOF}
	\end{figure}

	We begin by considering an individual spring-mass oscillator, as shown in \Figref{fig:SDOF}.
	This oscillator has mass~$ m $ and is attached, by an effective nonlinear spring with force $ f(t) $, to a moving base with prescribed displacement $ y(t) $. The displacement $ x(t) $ of the mass is measured with respect to its equilibrium position, \ie, when $ y(t) = 0 $ and in the absence of plastic deformation. Newton's Second Law yields the following first-order equations of motion of oscillator:
	\begin{equation}
		\label{eq:eomNewton}
		\begin{aligned}
			\dot{x} &= v, \\
			m\dot{v} &= f,
		\end{aligned}
	\end{equation}
	where $v$ is the velocity of the mass. 
	
	\subsection{Elastic-plastic Spring Element}

\begin{figure*}
			\includegraphics[width=1.0\linewidth]{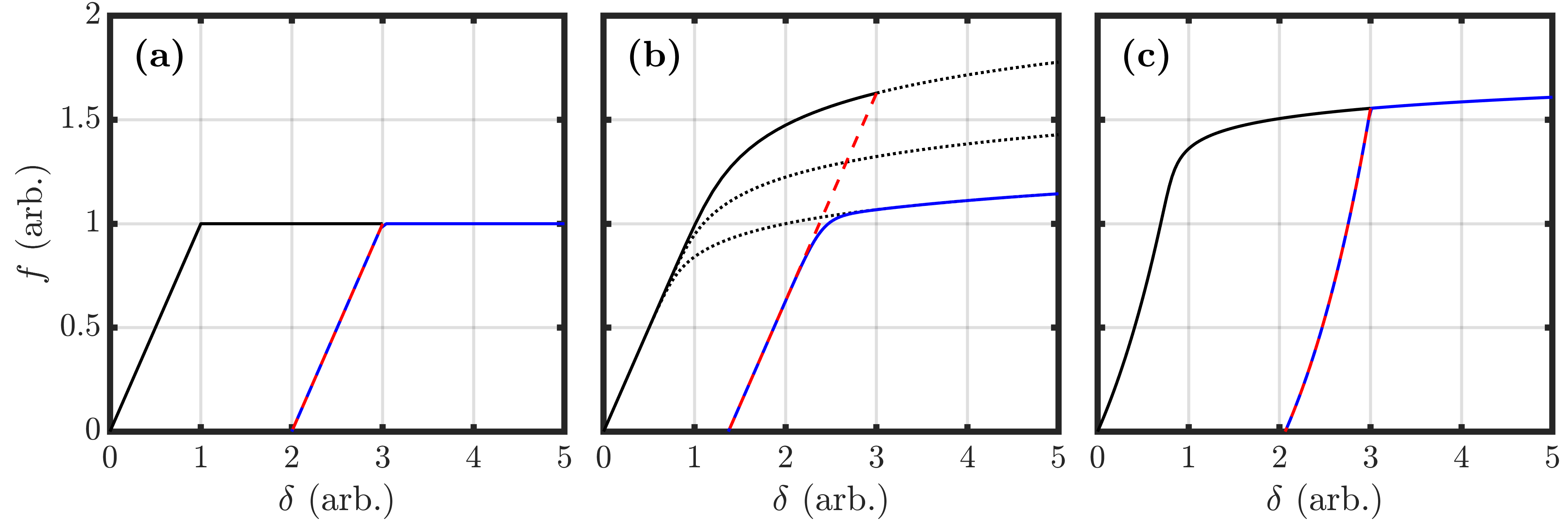}
			\caption{Force-displacement curves for three elasto-plastic effective spring models under displacement-controlled loading. Black solid, red dashed, and blue solid curves denote loading, unloading, and reloading portions, respectively. (a) Linear-elastic, perfectly-plastic model with $\ke=1$ and $\fy = 1$. (b) Rate-dependent power-law model with $\fy=1$, $\delta_0=0.1$, $\dot{\delta}_0=1$, $M = 10$, and $n = 0.1$. Loading and reloading occur at different prescribed rates, while unloading is assumed rate-independent. Black dotted curves correspond to the indicated rates. (c) Toda-Ramberg-Osgood model with $H=1.5$ and $n=0.05$.}
			\label{fig:PlasticityModels}
		\end{figure*}

		A typical lumped-parameter model for a one-dimensional, plastically deforming element, such as the one shown in \Figref{fig:SDOF}, comprises an elastic spring in series with a plastic slider \cite{DowlingBook}. The serial arrangement of the spring and slider is motivated by continuum plasticity theory, where the total strain may be decomposed additively into elastic (\ie, recoverable) and plastic (\ie, nonrecoverable) contributions \cite{NematNasser1979}. The total reduction in length $ \delta(t) $ of the effective spring element is thus the sum of elastic and plastic components of deformation, denoted $ e(t) $ and $ s(t) $, respectively, such that
		\begin{equation}\label{eq:RelDisp}
			\delta = y - x = e + s.
		\end{equation}
		As is evident from \Eqref{eq:RelDisp}, $ e(t) $ and $ s(t) $ take on positive values when the spring is \textit{compressed}. The two elements share a common force, which is related to their respective deformations (and, more generally, their rates of deformation) by constitutive relations describing the material response. Some plasticity models, such as the linear elastic, perfectly plastic (LEPP) model shown in \Figref{fig:PlasticityModels}(a)
		and the generalized Hertzian contact models discussed in \cite{Burgoyne2014,Burgoyne2015}, exhibit distinct values of force or deformation below which all deformation is elastic, and above which plastic deformation may occur. Conversely, other models admit plastic deformation regardless of loading (though it may be predominantly elastic at small loads \cite{nematnasser2004plasticity}), as is the case for the rate-dependent power-law model shown in \Figref{fig:PlasticityModels}(b). 
		We note the following significant differences between the model presented here and most DEM prevalent in the literature: i) the constitutive relationships may depend on the entire history of deformation and deformation rate, as opposed to the instantaneous values; ii) if the constitutive relations for the elastic and plastic elements may each be expressed in the form $f = g(e,\dot{e},s,\dot{s})$, then the resulting force equilibrium equation (\ie, equating the forces of the two elements in series) may be differential or algebraic; and iii) since the constitutive relationships are general nonlinear functions of the force and deformations, it may be impossible (or simply inconvenient) to formulate the equations of motion is terms of the minimal number of kinematic degrees-of-freedom. Thus, to accommodate a wide variety of constitutive relationships for elasto-plastic behavior, we consider general constraints of the form
		\begin{subequations}\label[Eqs]{eqs:constRels}
			\begin{align} 
				\gE(f,e,s,\xi) &= 0, \label{eq:constE}\\
				\gS(f,e,s,\xi) &= 0, \label{eq:constS}\\
				g_\xi(f,e,s,\xi) &= 0, \label{eq:constXI}
			\end{align}
		\end{subequations}
		where $\xi$ is an internal variable describing the history of deformation (such as a yield force or a characteristic of the NLEM microstructure), and, for ease of notation, we have suppressed the functional dependence on the rates $\edot $, $\sdot$, and $\xidot$. Additionally, the constitutive relations must be supplemented by a \textit{yield criterion}, which determines whether or not plastic deformation occurs. Mathematically, we state the yield criterion as 
		\begin{equation}\label{eq:yc}
			\phi_\mathrm{yc}(f,e,s,\xi) \ge 0,
		\end{equation}
		where we have again suppressed the functional dependence on time derivatives of the arguments. \Cref{eq:constE} describes the response of the elastic spring and is always enforced, while \Eqref{eq:constS,eq:constXI} are only enforced if \Eqref{eq:yc} satisfied. If \Eqref{eq:yc} is not satisfied, then the spring is in a purely elastic regime and the plastic variables $s$ and $\xi$ are held fixed. In the present study, we only consider yielding in compression; thus, \Eqref{eq:yc} is always supplemented by the condition $\edot + \sdot > 0$, regardless of the specific model.

	\subsection{Equations of Motion}\label{sec:EOMsingle}
		Given the formulation for the lumped-parameter model provided above, we write the complete equations of motion of the oscillator by assembling
		Eqs.~\eqref{eq:eomNewton}--\eqref{eq:yc}:
		\begin{subequations}\label[Eqs]{eq:eomSingle}
			\begin{align}
				0 &= \dot{x} - v, \\
				0 &= m\dot{v} - f,\\
				0 &= e + s + x - y,\\
				0 &= \gE(f,e,s,\xi),\\
				0 &= 
				\begin{cases}			
					\dot{s} & \text{if } \phi_\mathrm{yc}(f,e,s,\xi) < 0	\\
					\gS(f,s,e,\xi)  & \text{if } \phi_\mathrm{yc}(f,e,s,\xi) \ge 0
				\end{cases}, \label{eq:eomS}\\
				0 &= 
				\begin{cases}			
					\xidot & \text{if } \phi_\mathrm{yc}(f,e,s,\xi) < 0	\\
					g_\xi(f,s,e,\xi) & \text{if } \phi_\mathrm{yc}(f,e,s,\xi) \ge 0
				\end{cases}, \label{eq:eomXI} 
			\end{align}
		\end{subequations}
		where, for \Eqref{eq:eomS,eq:eomXI}, the (top, bottom) equation is enforced when the spring is in an (elastic, plastic) regime. In general, \Cref{eq:eomSingle} comprise a system of six differential-algebraic equations (DAEs) in the six unknowns $x$, $v$, $f$, $e$, $s$, and $\xi$. Details about the numerical solution of DAEs in this work are presented in \Appref{app:DAE}.

	\subsection{Example: Linear Elastic/Perfectly Plastic}\label{sec:LEPPsingle}
	
		As a first illustration of the single-oscillator framework, we consider the LEPP model, which consists of a linear elastic regime with stiffness $k_e$ and damping coefficient $c_e$, followed by a plastic regime with constant force $f_\yy$. The constitutive relations for this model may be expressed in the form of \cref{eqs:constRels} as follows:
		\begin{subequations}\label[Eqs]{eq:crPP}
			\begin{align}
				\gE &= f - \ke e - \ce \edot = 0, \\
				\gS &= f - \xi = 0, \\
				g_\xi &= \xi - f_\yy = 0,
			\end{align}
		\end{subequations}
		where we have chosen the yield force $f_\yy$ as the internal variable $\xi$. Similarly, the yield criterion is given by
		\begin{equation}
			\phi_\yc = f - \xi \ge 0.
		\end{equation}
		While $\xi$ may seem redundant for the LEPP model, this form generalizes straightforwardly to more complex plasticity models for which the yield force depends on the deformation history.
		
\begin{figure}
			\includegraphics[width=1.0\linewidth]{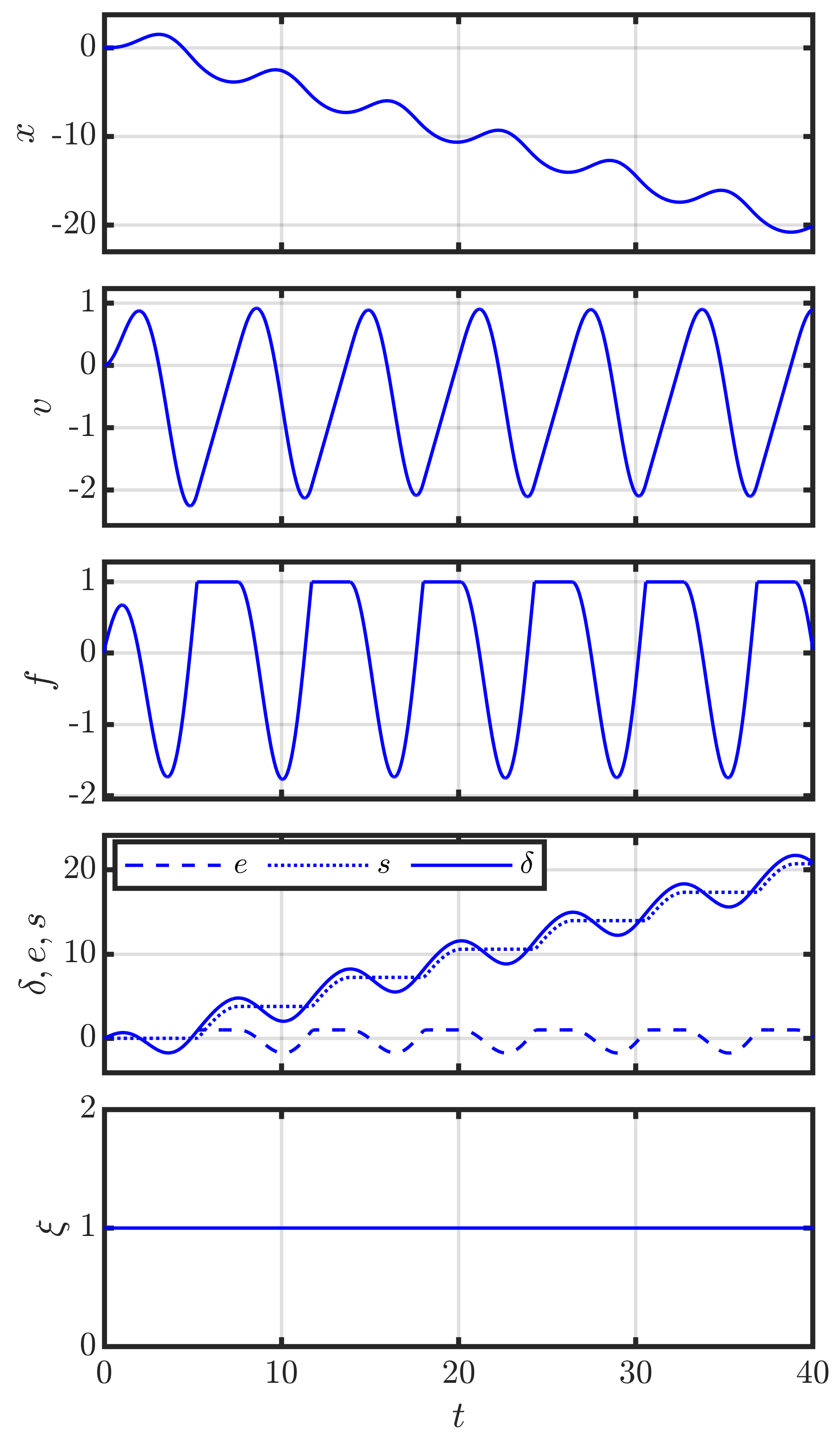}
			\caption{Numerical solution of a spring-mass oscillator with a LEPP constitutive relationship and time-harmonic forcing near the linear resonance. Panels contain time histories of (a) displacement and (b) velocity, respectively, of the mass; (c) force in spring and slider; (d) deformations of the spring and slider elements, as well as the total deformation; and (e) internal variable $\xi$, which stores the value of the constant yield force $\fy$ for the LEPP model.}
			\label{fig:singleoscpp-v1}
		\end{figure}

		Using the parameters $m=\ke=f_\yy = 1$, $\ce=0.05$, and the excitation $y(t) = \sin(t)$ (all with arbitrary, dimensionless units), we numerically integrate \cref{eq:eomSingle} via the 3-stage Radau-IIA method with a time step $h = 0.01 $. The solution is shown in \Figref{fig:singleoscpp-v1}. 
		Since the oscillator is excited by a time-harmonic boundary displacement near its linear natural frequency, it undergoes a resonant response until the force reaches the yield value $f_\yy=1$. After a few cycles of loading, it reaches a steady-state response in which the equilibrium displacement shifts downward due to successive yielding with each cycle.

\section{Mass-Spring Chains}\label{sec:Chains}

	\begin{figure*}
		\includegraphics[width=1.0\linewidth]{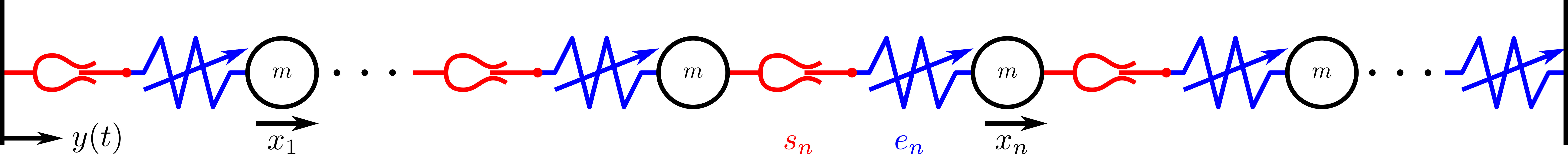}
		\caption{Schematic of a chain of coupled elasto-plastic oscillators. All symbols and colors are analogous to the individual oscillator shown in \Figref{fig:SDOF}.}
		\label{fig:plasticchainschematic}
	\end{figure*}
	
	Having developed the equations of motion for a single oscillator in \Secref{sec:EOMsingle}, we now extend the model to a chain of $N$ masses connected by elastic-plastic spring elements, as shown in \Figref{fig:plasticchainschematic}. We begin by writing the equations of motion of an arbitrary interior unit, and then handle the units on the boundaries. For the interior unit with index $n$, the equations of motion are the following:
	\begin{subequations}\label[Eqs]{eq:eomChain}
		\begin{align}
			0 &= \dot{x}_n - v_n, \label{eq:eomXn}\\
			0 &= m\dot{v}_n - f_n + f_{n+1}, \label{eq:eomVn}\\
			0 &= e_n + s_n + x_n - x_{n-1},\label{eq:eomEn}\\
			0 &= \gE(f_n,e_n,s_n,\xi_n),\\
			0 &= 
			\begin{cases}			
				\dot{s}_n & \text{if } \phi_\mathrm{yc}(f_n,e_n,s_n,\xi_n) < 0	\\
				\gS(f_n,e_n,s_n,\xi_n)  & \text{if } \phi_\mathrm{yc}(f_n,e_n,s_n,\xi_n) \ge 0
			\end{cases}, \label{eq:eomSn}\\
			0 &= 
			\begin{cases}			
				\xidot_n & \text{if } \phi_\mathrm{yc}(f_n,e_n,s_n,\xi_n) < 0	\\
				g_\xi(f_n,e_n,s_n,\xi_n) & \text{if } \phi_\mathrm{yc}(f_n,e_n,s_n,\xi_n) \ge 0
			\end{cases}. \label{eq:eomXIn} 
		\end{align}
	\end{subequations}
	The only differences between \cref{eq:eomChain} and \cref{eq:eomSingle}, other than including the index $n$, are i) the force due to the spring element to the right of mass $n$ is included in \Eqref{eq:eomVn}; and ii) the prescribed boundary displacement $y(t)$ in \Eqref{eq:eomEn} is replaced by the displacement of the mass to the left. For the mass adjacent to the left boundary (i.e., $n=1$), only \Eqref{eq:eomEn} need be modified:
	\begin{equation}
		0 = e_1 + s_1 + x_1 - y(t). \label{eq:eomE1}
	\end{equation}
	For the present study, we couple the right-most mass (i.e., $n=N$) to a rigid boundary by a linear, elastic spring with stiffness $k_b$, where, for each model, $k_b$ is the slope of the linearized force-displacement curve at the origin. This is implemented in the equations of motion by replacing \Eqref{eq:eomVn} with the following:
	\begin{equation}
		0 = m\dot{v}_N - f_N + k_b x_N.
	\end{equation}

	In the following subsections, we solve \cref{eq:eomChain} for a chain of length $N=21$ masses, using three elasto-plastic models with distinctly different behaviors and mathematical forms. For each case, we apply a smooth, step-like excitation of the form
	\begin{equation}\label{eq:stepLike}
		y(t) = \begin{cases}
			0 & \text{if } t < 0, \\
			\frac{A}{2}\left[ 1 - \cos{\left(\frac{\pi t}{w}\right)}\right] & \text{if } 0 \le t \le w,\\
			A & \text{if } t > w,
		\end{cases}
	\end{equation}
	where $A$ is the step amplitude and $w$ is the width. Since these examples are intended to illustrate the DEM framework in the general sense, all units are arbitrary and dimensionless.
	
	\subsection{Linear Elastic/Perfectly Plastic}

		To establish a baseline for elasto-plastic chain models, we first simulate a linear, conservative chain and examine the results when perfectly-plastic yielding (as defined earlier in \Secref{sec:LEPPsingle}) is introduced. Using the parameters $A=4$ and $w=4$ in the step-like excitation defined by \Eqref{eq:stepLike}, we realize both models via the constitutive relations given by \Eqref{eq:crPP}, with $\ke=1$, $\ce=0$, and $\fy = (2,1)$ for the linear and LEPP cases, respectively (\ie, for the linear case, the force never exceeds the yield value and the dynamics are equivalent to a classical mass-spring chain). 
		\begin{figure*}
			\includegraphics[width=1.0\linewidth]{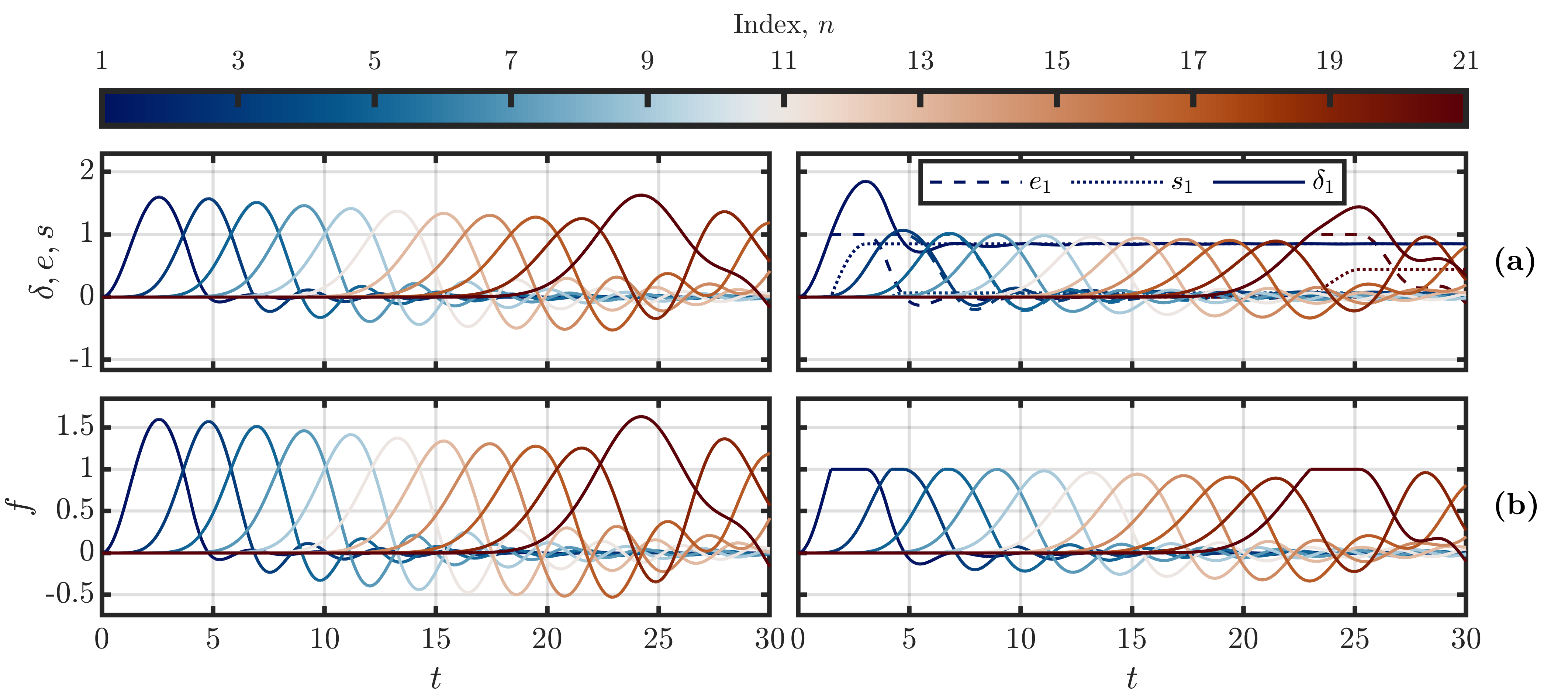}
			\caption{Numerical solution of a 21-mass chain with LEPP constitutive relationship and step-like excitation at the left boundary. Rows contain time histories of (a) deformation (b) force. (Left, right) columns contain data for which the yield force $\fy$ is (greater than, equal to) the maximum force developed during the simulation.}
			\label{fig:chainPP}
		\end{figure*}
		As shown in the left-hand column of \Figref{fig:chainPP}, in the absence of yielding, the propagation is linear and dispersive with the longest-wavelength disturbances propagating with greatest speed (see, \eg, \cite{BrillouinBook} for a detailed treatment of this classical model). When yielding does occur, as shown in the right-hand column of \Figref{fig:chainPP}, the yield force $\fy$ is the maximum value of compressive force that may be transmitted in the chain; plastic deformation occurs in the first few units, after which the propagation is linear and dispersive.
		
	\subsection{Rate-Dependent Power Law}
		Next we examine a discrete analogue of a rate-dependent power law (RDPL) model described in \cite{nematnasser2004plasticity}, where the rate of plastic displacement is defined by the following relationship \footnote{See Eq.~(4.8.2) in \cite{nematnasser2004plasticity}, which uses different variable names.}:
		\begin{equation}
			\frac{\sdot}{\sdot_0} = \left[\frac{f}{f_0\left(1+s/s_0\right)^\nu}\right]^\mu,
		\end{equation}
		where $f_0$, $s_0$, and $\sdot_0$ are reference values of the force, plastic displacement, and rate of plastic displacement, respectively, and $\mu$ and $\nu$ are dimensionless parameters that determine the shape of the hardening curve. The RDPL model does not have a distinct yield point; that is, any positive force generates a plastic displacement, though the total displacement may be dominated by the elastic contribution at low values of force. Using a linear relationship for the elastic displacement, and the internal history parameter \footnote{While $\xi$ is not needed to fully define the model, this definition yields less cumbersome expressions in the Jacobian matrix used for numerical integration.} $\xi = (1+s/s_0)^\nu$, the constraints and yield criterion for the DEM framework may be written
		\begin{subequations}
			\begin{align}
				\gE &= f - \ke e  = 0, \\
				\gS &= \sdot - \sdot_0\left(\frac{f}{f_0 \xi}\right)^\mu  = 0, \\
				g_\xi &= \xi - \left(1+s/s_0\right)^\nu = 0, \\
				\phi_\yc &= f \ge 0.
			\end{align}
		\end{subequations}

\begin{figure*}
			\includegraphics[width=1.0\linewidth]{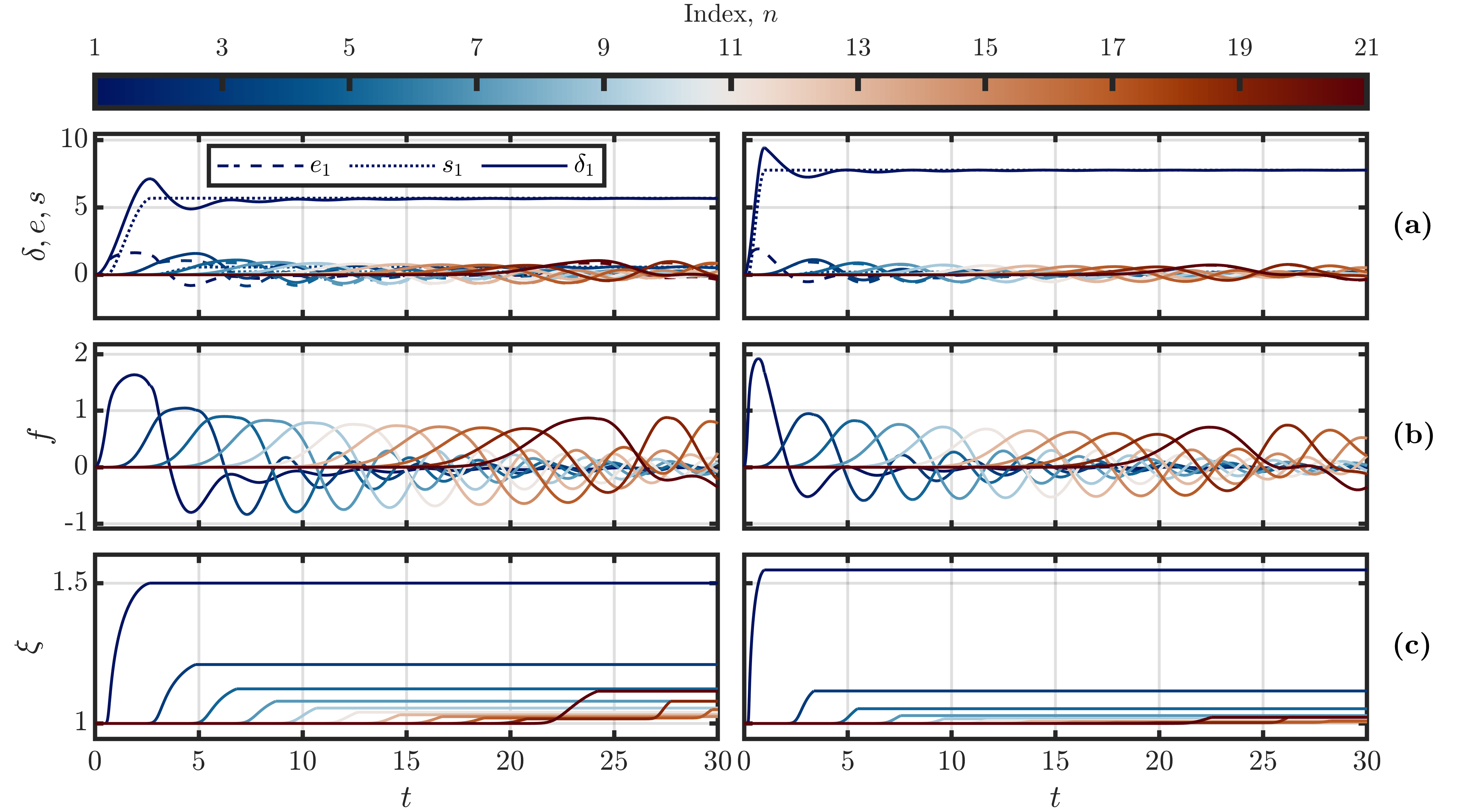}
			\caption{Numerical solution of a 21-mass chain with RDPL constitutive relationship and step-like excitation at the left boundary. Rows contain time histories of (a) deformation, (b) force, and (c) internal variable $\xi$. (Left, right) columns contain data for step-like excitations with average loading rates $A/w=$ (3.33, 10).}
			\label{fig:chainPL}
		\end{figure*}

		For numerical simulations of the RDPL model, we select the parameters $f_0=1$, $s_0 = 0.1$, $\sdot_0=1$, $\mu = 10$, and $\nu=1/10$. Example force-displacement curves for a displacement-controlled loading sequence are shown in \Figref{fig:PlasticityModels}(b), where the prescribed rate of total displacement is piecewise-constant, with different values for the loading and reloading segments. To observe rate-dependent behavior, we apply step-like excitations with amplitude $A=10$ and two widths: $w=3$ and $w=1$. We note that, for each of these cases, the average loading rate $A/w$ is significantly greater than the long-wavelength sound speed of the linearized chain, $\sqrt{\ke/m} = 1$~(unit cells/time). The numerical solutions for these two cases are shown in the left- and right-hand columns of \Figref{fig:chainPL}, respectively. Comparing these two simulations, we observe that the higher loading rate results in higher peak force and plastic displacement in the first unit cell, but lower peak values transmitted through the chain. However, as shown in \Figref{fig:chainPL}(c), the lattice sites beyond the first experience greater plastic displacement for the lower loading rate.

	\subsection{Plastic Toda Lattice}
		As a final example, we consider an elasto-plastic extension of the so-called Toda lattice \cite{Toda1970}, which is a completely-integrable, 1D lattice model supporting exact soliton solutions (\ie, localized traveling wave solutions that propagate without distortion, due to interplay between nonlinearity and dispersion) \cite{Shen2014}. The elastic deformation is related to the force via an exponential relationship:
		\begin{equation}
			f = \exp(e)-1.
		\end{equation}
		We incorporate plascitiy into the Toda lattice via the Ramberg-Osgood plasticity model \cite{ROreport,DowlingBook}, for which the plastic deformation is related to the force via the relation
		\begin{equation}
			s = \left(\frac{f}{f_0}\right)^{1/\nu},
		\end{equation}
		where $f_0$ is a reference value of force and $\nu$ is a dimensionless parameter that determines the shape of the hardening curve. A representative loading sequence for the Toda-Ramberg-Osgood (TRO) model is shown in \Figref{fig:PlasticityModels}(c). This model does not have a distinct yield point in the first loading cycle; however, unloading and reloading are elastic until the prior maximum value of force is reached. Thus, to track this variable yield force, we choose the force as the internal variable $\xi$. The complete constitutive relations of the TRO model may be written in the DEM framework as follows:
		\begin{subequations}\label[Eqs]{eq:constRelTRO}
			\begin{align}
				\gE &= f - \exp{(e)} + 1  = 0, \\
				\gS &= s - \left(\frac{f}{f_0}\right)^{1/\nu}  = 0, \\
				g_\xi &= \xi - f = 0, \\
				\phi_\yc &= f - \xi \ge 0.
			\end{align}
		\end{subequations}
		
		For numerical simulations of the TRO model, we select the model parameters $f_0=1.5$ and $\nu=1/20$ and use the amplitude $A=3$ and width $w=2$ for the step-like excitation. To elucidate the effects of plasticity, we also simulate a Toda lattice without plasticity. 
		\begin{figure*}
			\includegraphics[width=1.0\linewidth]{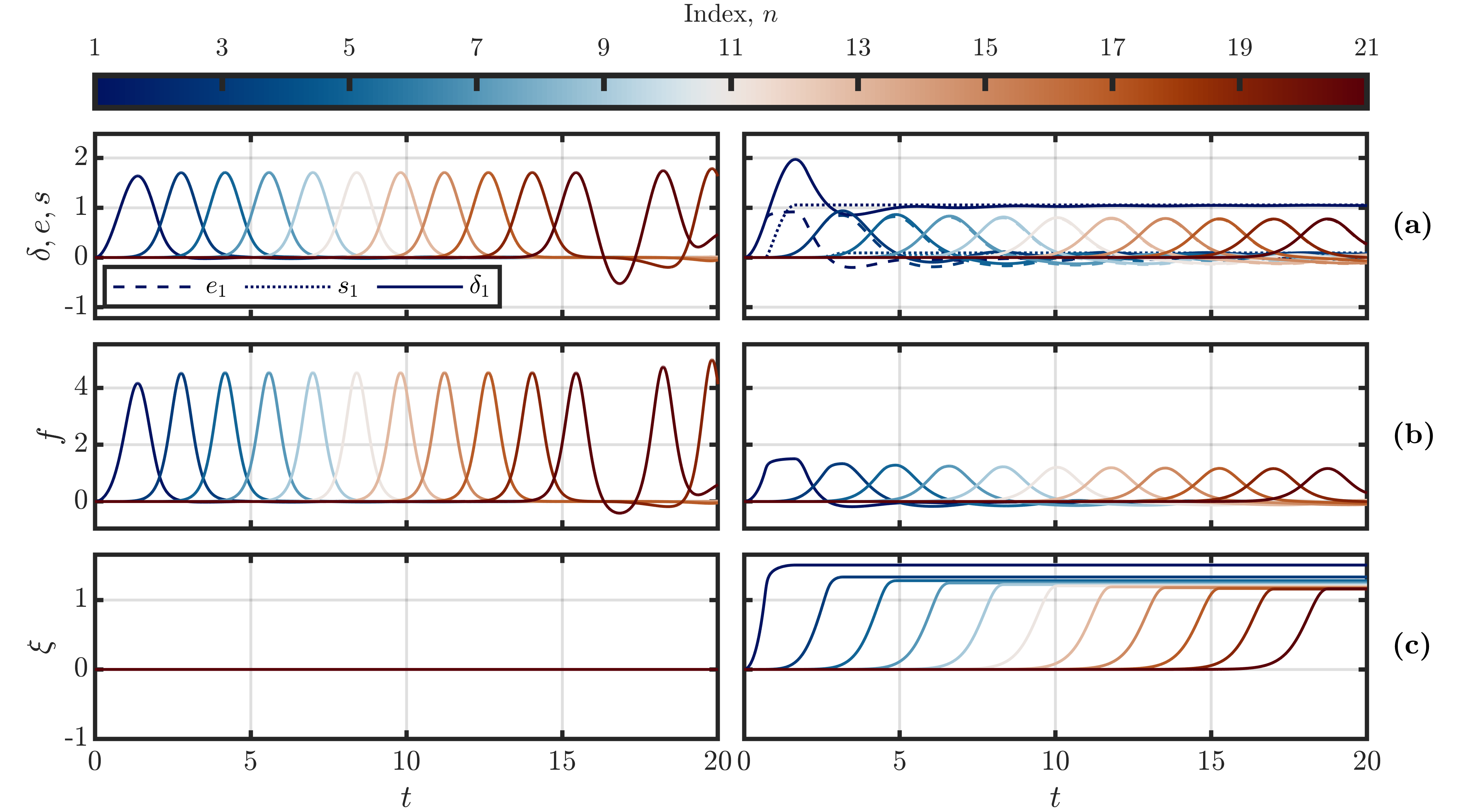}
			\caption{Numerical solution of a 21-mass chain with TRO constitutive relationship and step-like excitation at the left boundary. Rows contain time histories of (a) deformation, (b) force, and (c) internal variable $\xi$. (Left, right) columns contain data for Toda lattices (without, with) plasticity.}
			\label{fig:chainTRO}
		\end{figure*}
		\Figref{fig:chainTRO} shows that, when the lattice is assumed to be purely elastic, the initial pulse evolves into a soliton and propagates without distortion until interacting with the opposite boundary. For the case with plasticity included, the first few lattice sites undergo significant yielding until an approximate (but dissipative) soliton forms with a reduced peak force value.

\section{Simulated Buckling Element}\label{sec:NSH}
	
	For general lattice geometries undergoing finite deformation (see, \eg, \cite{Kochmann2017rev,Liu2024review} for numerous examples), simple analytical constitutive models in the plastic regime (such as those discussed in \Secref{sec:Chains}) are generally not available. In this section, we demonstrate a procedure to obtain empirical constitutive relationships from simulated or experimental mechanical test data. For the present study, we develop the procedure in the context of a negative-stiffness honeycomb (NSH) geometry, which is designed to buckle at a prescribed threshold force value in order to mitigate accelerations associated with mechanical impacts \cite{Correa2015,Goldsberry2018}. A representative schematic of a chain of NSH elements, which we aim to model, is shown in \Figref{fig:nshfeageomv1}(a). While some portions of the analysis are necessarily geared toward our particular data set, the general strategy need not be restricted to this specific geometry. The procedure is outlined as follows:
	\begin{enumerate}
		\item Obtain, via simulation or experiment, a data set containing cyclical force-displacement data, with progressively increasing levels, for a representative sample of the material under study. Partition the data into loading and unloading segments.
		\item Fit a single, continuous curve to the all of the loading data.
		\item Fit a continuous curve to each segment of unloading data.
		\item Parameterize the fitted curves in terms of the plastic deformation $s$, and, if needed, some history-dependent internal variable $\xi$.
	\end{enumerate}
	For the present study, we obtain a simulated data set via finite element analysis (FEA). We discuss each step in detail in the following subsections.
	
	\subsection{Simulated Data Set via Finite Element Analysis}

\begin{figure}
			\includegraphics[width=1.0\linewidth]{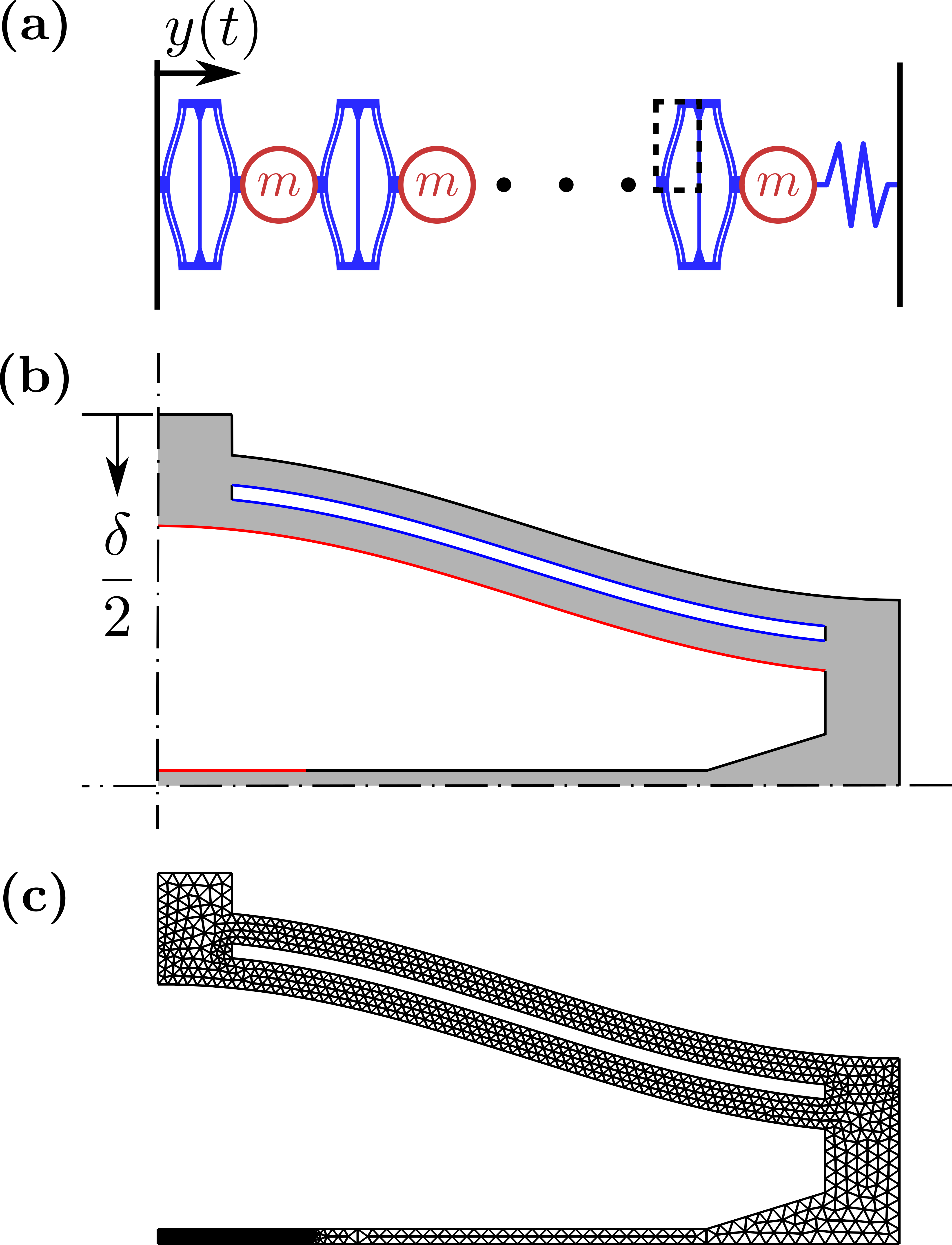}
			\caption{(a) Schematic of spring-mass chain composed of NSH elements, excited by boundary displacement $y(t)$. The black dashed rectangle encloses the region modeled using FEA (rotated $90^\circ$ counter-clockwise with respect to the geometry shown in (b)-(c)).
			(b) Computational geometry for FEA of a NSH element. The following boundary conditions are applied: symmetry (black dash-dotted), contact pairs (blue and red solid), prescribed displacement $\delta/2$ (top surface), and traction free (all others). (c) Finite element mesh. The very dense portion in the lower left-hand corner is required for convergence if the red contact pair shown in (b) becomes active.}
			\label{fig:nshfeageomv1}
		\end{figure}

		To obtain our simulated data set, we perform FEA using the \comsol\ software package with the Nonlinear Structural Mechanics Module. The computational geometry, which is two-dimensional with plane-strain conditions assumed, is shown in \Figref{fig:nshfeageomv1}(b), with geometric parameters listed in Table~\ref{tab:NSH}.
          \begin{table}
        		\caption{\label{tab:NSH} Geometric parameter values for NSH geometry. All parameters are defined as in \cite{Goldsberry2018}, though the numerical values differ from that work. All values are given in units of mm.}
        		\begin{ruledtabular}
        			\begin{tabular}{lcd}
        				Parameter & Description & \text{Value} \\
        				\hline
        				$L_x$ & Horizontal length & 50.0\\
        				$L_y$ & Vertical length & 30.0\\
        				$t_b$ & Beam thickness & 1.0\\
        				$t_s$ & Beam separation & 0.5\\
        				$h_b$ & Beam apex height & 5.0\\
        				$h_c$ & Center height & 2.0\\
        				$w_c$ & Center width & 4.0\\
        				$h_{cb}$ & Center beam height & 1.25\\
        				$w_{cb}$ & Center beam width & 5.0\\
        				$t_{hb}$ & Horizontal beam thickness & 1.0 
        			\end{tabular}
        		\end{ruledtabular}
        	\end{table}
		We select material properties to approximate 6061-T6 aluminum with Young's modulus $E=68$~GPa and Poisson's ratio $\nu=0.33$, with finite elastic deformation included via the St.~Venant-Kirchoff hyperelastic model. Material plasticity is incorporated via the Johnson-Cook \cite{JohnsonCook1985} model \footnote{
		The classical Johnson-Cook model contains a rate-dependent, logarithmic term that grows without bound in the limit of vanishing rate of equivalent plastic strain. For quasi-static problems, such as the present case, \comsol\ neglects this term (or, equivalently, assumes that the rate of equivalent plastic strain is equal to the nominal value present as a model parameter, such that the argument of the logarithm takes a value of 1). We also neglect thermal softening, which may be captured by the Johnson-Cook model in principle.
		}
		with the following parameters: \footnote{
		The following symbols are chosen for consistency with the \comsol\ interface; some of them are duplicate variables within the present work.
		} 
		yield stress $\sigma_\mathrm{ys0}=270$~MPa, strength coefficient $k = 165.5$~MPa, and hardening exponent $n=0.222$ \cite{Giglio2014}. We apply cyclic, displacement-controlled loading $\delta/2$ in increments of 0.01~mm and increase the maximum displacement by 0.03~mm with each cycle. For each value of $\delta$, we compute the effective spring force $f$ by measuring the total reaction force on the controlled boundary. To prevent reverse-yielding, the unloading portion of each cycle is terminated when the total reaction force becomes negative. To account for possible self-contact under large deformation, contact pairs are assigned. However, for the results herein, we restrict our analysis to the data prior to the first instance of self-contact. Finally, to reduce computational cost, we apply symmetry boundary conditions along the relevant axes. The finite element mesh is shown in \Figref{fig:nshfeageomv1}(c).
	
\begin{figure*}
			\includegraphics[width=0.8\linewidth]{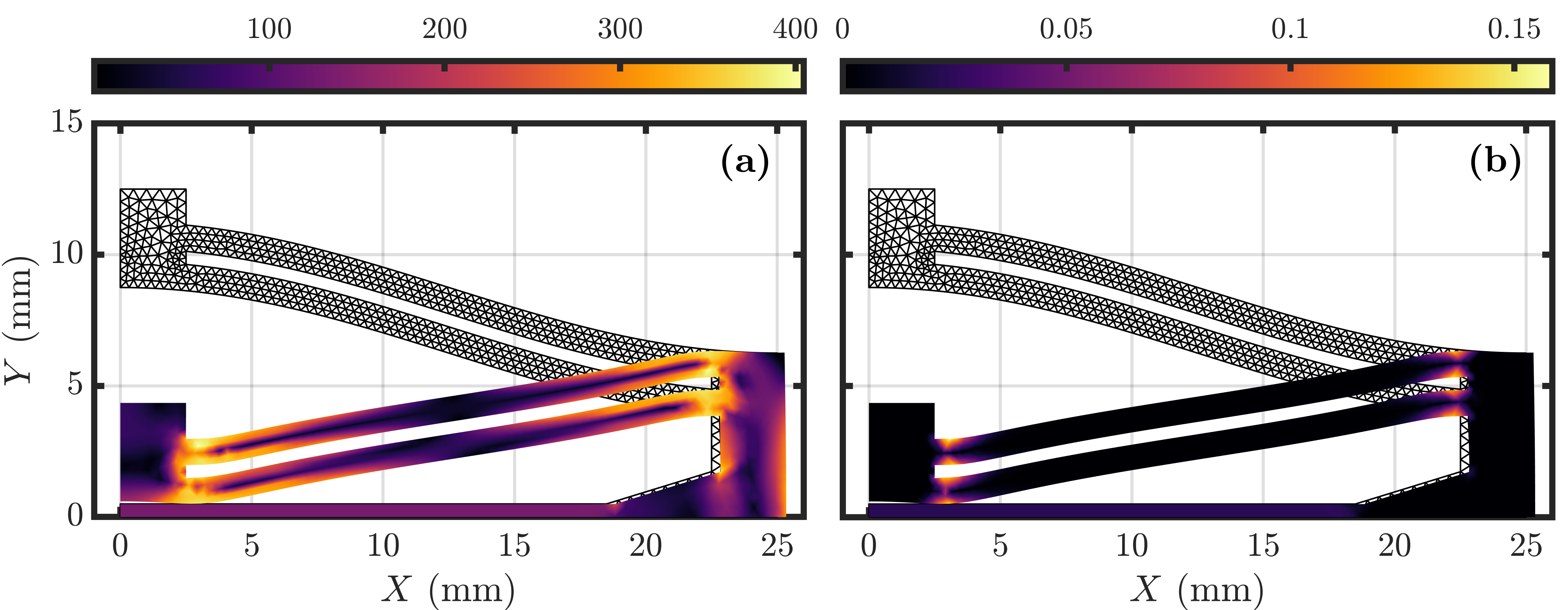}
			\caption{Surface plots of (a) von Mises stress (MPa), and (b) equivalent plastic strain (dimensionless), from FEA of a NSH element, immediately prior to the initiation of self-contact.}
			\label{fig:nshplastic-feafieldsv1}
		\end{figure*}

\begin{figure}
			\includegraphics[width=0.8\linewidth]{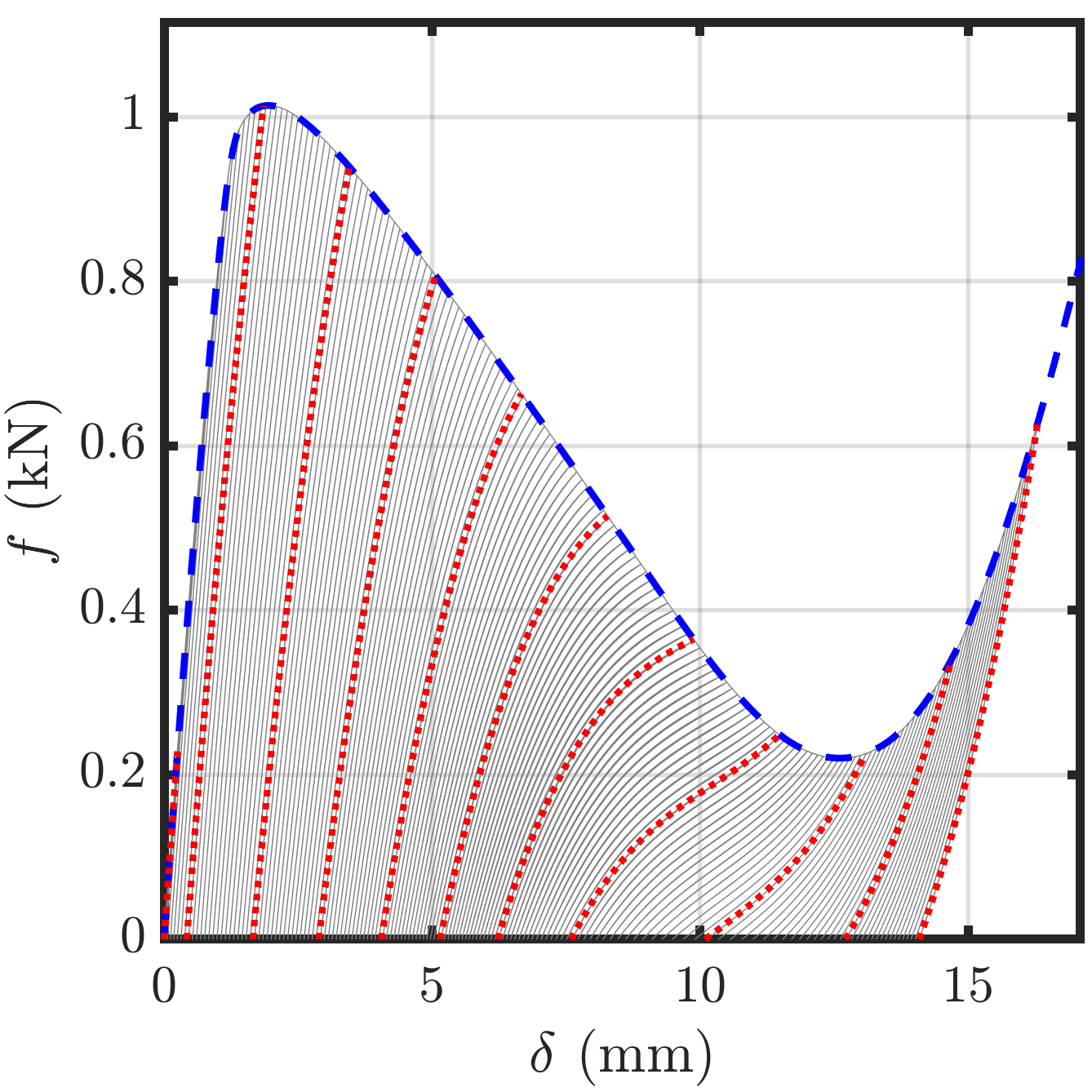}
			\caption{Force-deformation relationship obtained from FEA of a NSH unit cell. Thin gray curves: synthetic data obtained from FEA. Dashed blue curve: piecewise cubic spline fitted to loading segments of the FEA data. Dotted red curves: cubic Hermite splines fitted to unloading segments of the FEA data, generated at several arbitrary points on the loading curve.}
			\label{fig:nshCurve}
		\end{figure}

		The computed von Mises stress and equivalent plastic strain fields, for a solution immediately prior to the onset of self-contact, are shown in \Figref{fig:nshplastic-feafieldsv1},
		and the force-deformation data for the entire simulation run is shown as thin, gray curves in \Figref{fig:nshCurve}.
		As shown in \Figref{fig:nshplastic-feafieldsv1}, the stress and strain are generally greatest near the attachment points of the curved, beam-like portions of the geometry, with the plastic strain localized in particularly small zones. While these plastic zones occupy a very small fraction of the total material volume, the macroscopic displacement $\delta$ is predominantly plastic, as shown by the right-most gray curve in \Figref{fig:nshCurve}.

	\subsection{Empirical Elasto-Plastic Model}
		To develop a discrete spring model for the NSH from the FEA data, we first fit a curve to the loading portions of the force-deformation data. This curve is not well-captured by a single polynomial; thus, we fit it using a piecewise cubic spline \footnote{For experimentally-obtained data, the data would first need to be smoothed.}, as shown by the dashed, blue curve in \Figref{fig:nshCurve}. 
		
\begin{figure}
			\includegraphics[width=0.8\linewidth]{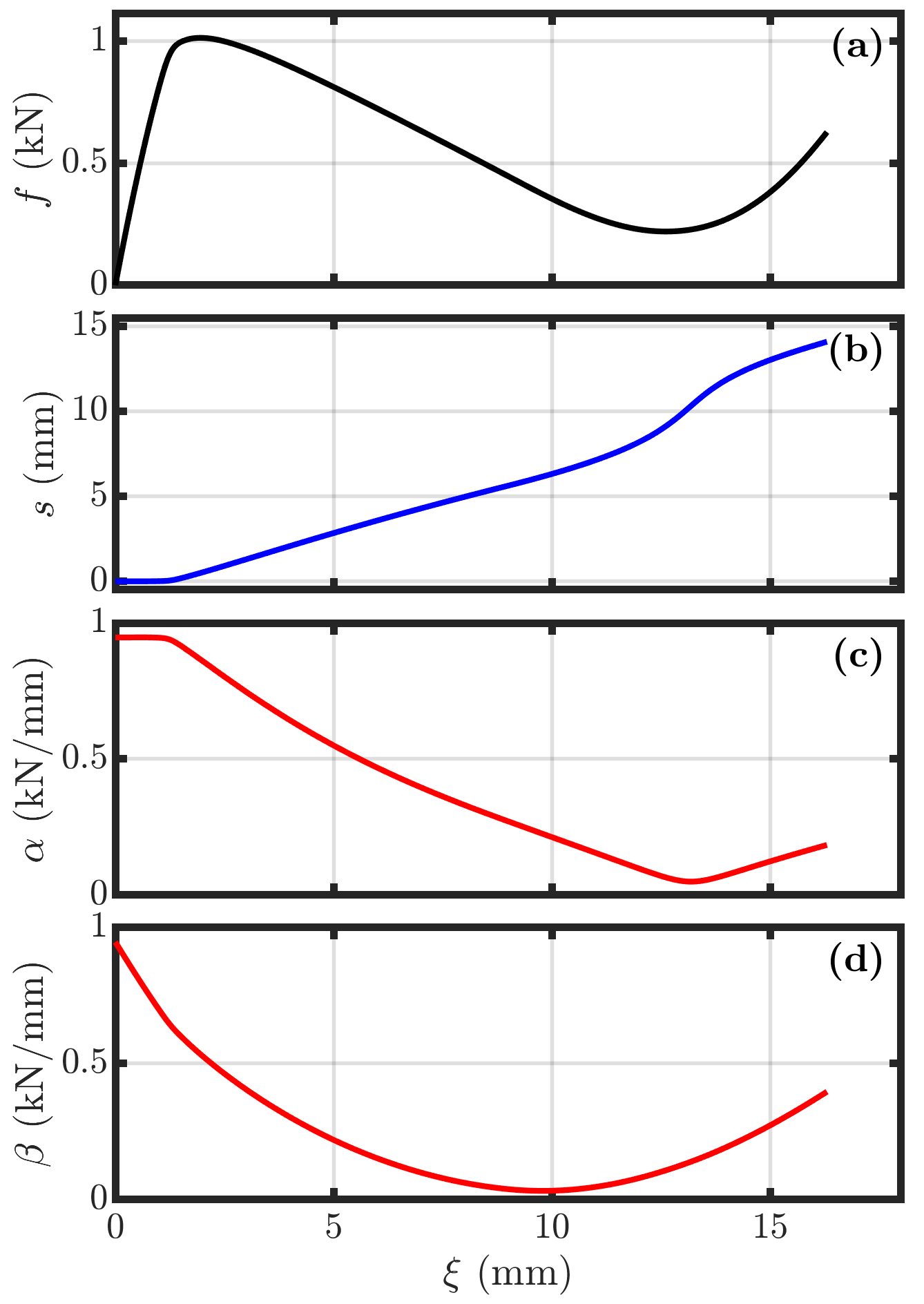}
			\caption{Fitted parameters used to define cubic Hermite splines interpolating elastic unload/reload curves, for an arbitrary maximum displacement $\xi$. (a) Force prior to unloading. (b) Plastic displacement after unloading, i.e., when $f=0$. (c) Slope of unloading curve at zero force. (d) Slope of the unloading curve at maximum displacement (\ie, the onset of unloading).}
			\label{fig:nshcurvefitcoeffsv1}
		\end{figure}

		Before fitting curves to the unloading data, the following points are in order: i) the model must have the ability to generate, from an arbitrary point on the loading curve, an unloading curve that exactly intersects the loading curve at that point; and ii) each unloading curve must capture the measured plastic deformation at zero force. In other words, the critical information contained in each unloading curve is defined by its endpoints. A suitable function to represent each unloading curve is therefore the cubic Hermite spline \cite{Burden1997Book} defined by
		\begin{multline}
			f(\delta) = \left[h_{00}(\theta)f_1 + h_{01}(z)f_2 \right. \\ 
			          + \left. h_{10}(\theta)(\delta_2-\delta_1)f'_1 + h_{11}(\theta)(\delta_2-\delta_1)f'_2\right],
		\end{multline}
		where $f_{1,2}$ and $\delta_{1,2}$ are the force and deformation at each endpoint of the curve, $f'_{1,2}$ is the corresponding slope, the functions $h_{ij}$ are the cubic Hermite basis functions, and $\theta=(\delta-\delta_1)/(\delta_2-\delta_1)$ is a shifted and scaled deformation. Noting that the force $f_1$ at the left endpoint is zero by construction, we determine the other coefficients for each unloading segment by performing a least-squares cubic polynomial fit, numerically locating the nearest root of the fitted polynomial (\ie, $\delta_1$) and the intersection with the loading curve ($\delta_2$ and $f_2$), and evaluating the derivative of the fitted polynomial at these two points ($f'_1$ and $f'_2$). After performing these calculations for every segment of unloading data, we fit each coefficient as a function of the maximum deformation, $\delta_2$, using piecewise cubic splines. Thus, a Hermite spline can be completely defined at any point on the loading curve. Finally, using the maximum deformation $\delta_2$ as the internal variable $\xi$ and noting that $\delta_1=s$ and $\delta-\delta_1=e$, the entire empirical model may be defined in the DEM framework as follows:
		\begin{subalign}[eq:constRelNSH]
			g_e &= f - \left[h_{01}(\theta)F(\xi) + h_{10}(\theta)(\xi-s)\alpha(\xi)\right. \nonumber \\
			&\qquad+ \left. h_{11}(\theta)(\xi-s)\beta(\xi)\right] = 0,\label{eq:crNSHe}\\
			g_s &= s - S(\xi) = 0,\label{eq:crNSHs} \\
			g_{\xi} &= \xi - (e+s) = 0,\\
			\phi_\yc &= \xi - (e+s) \ge 0,
		\end{subalign}
		where $\theta = e/(\xi-s)$, and $F(\xi)$, $S(\xi)$, $\alpha(\xi)$, and $\beta(\xi)$ are the piecewise cubic splines defining the loading curve, plastic offset, and slopes at the left and right end points, respectively, as functions of the maximum deformation $\xi$. The four piecewise cubic splines are shown in \Figref{fig:nshcurvefitcoeffsv1},
		and Hermite splines are shown for several representative values of $\xi$ in \Figref{fig:nshCurve}.

	\subsection{DEM Simulation}
\begin{figure*}
			\includegraphics[width=1.0\linewidth]{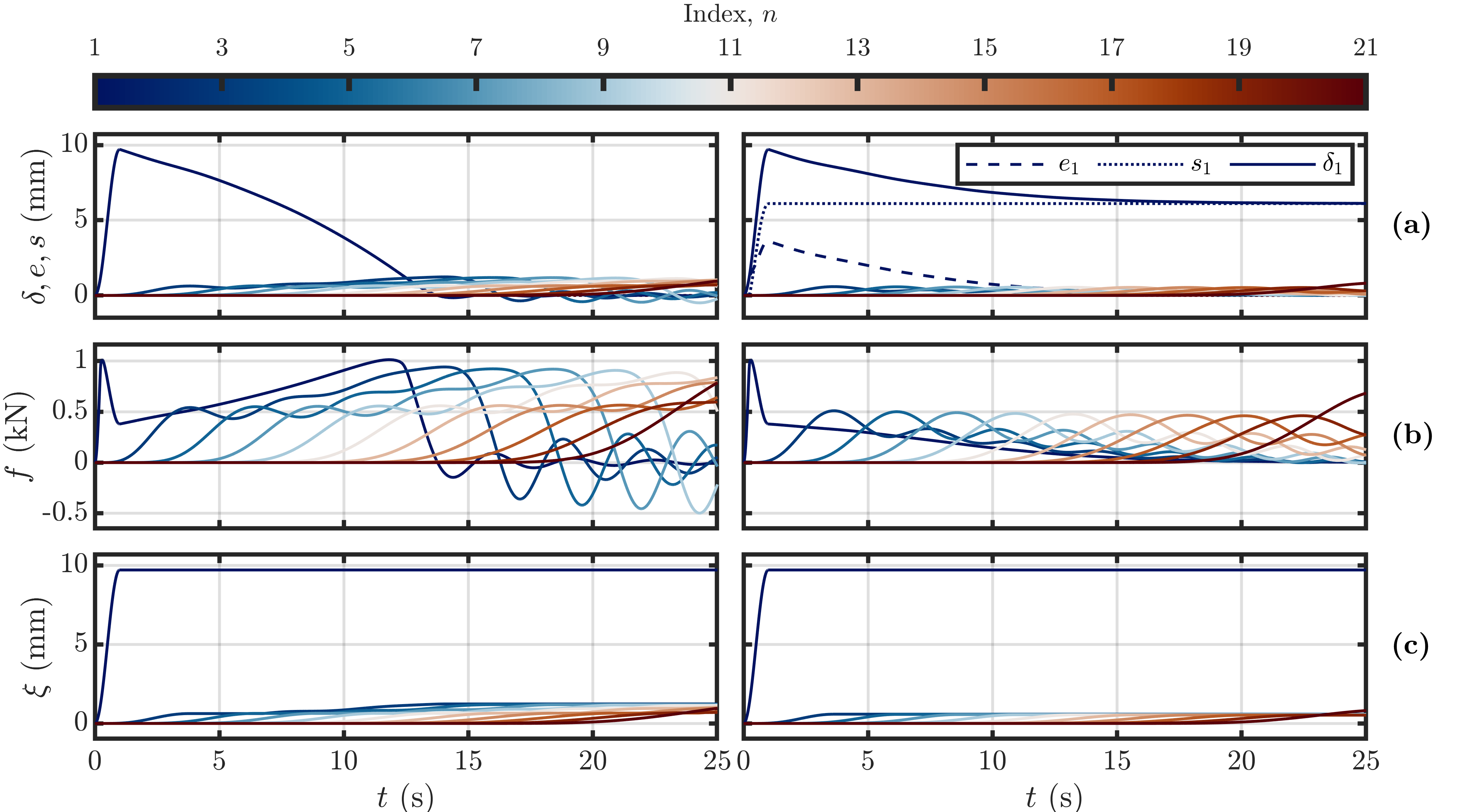}
			\caption{Numerical solution of a 21-mass chain with empirical constitutive relationship derived from FEA of a NSH unit, with step-like excitation at the left boundary. Rows contain time histories of (a)-(b) displacement and velocity, respectively, of the mass; (c) deformation of the spring and slider elements; (d) force in the elastic spring and plastic slider; and (e) internal variable $\xi$, which tracks the maximum total deformation developed in each element. (Left, right) columns contain data for lattices (without, with) plasticity.}
			\label{fig:plasticchainnshv1}
		\end{figure*}

		Finally, we compute numerical solutions of the DEM using the fitted NSH model defined by \Cref{eq:constRelNSH}, with mass $m=1$\units{kg} and the step-like excitation parameters $A=5$\units{cm} and $w=1$\units{s}. To elucidate the effects of plasticity, we repeat this computation using a modified NSH model in which \cref{eq:crNSHe,eq:crNSHs} are replaced by
		\begin{equation}
			g_e = f - F(e) = 0
		\end{equation} 
		and
		\begin{equation}
			g_s = s - \sdot = 0,
		\end{equation} 
		respectively, simulating a naive compression test in which the entire force-deformation curve is measured in the loading direction only and assumed elastic (the maximum deformation is still stored in $\xi$ but does not influence the model behavior). As shown in \Figref{fig:plasticchainnshv1},
		comparing the left- and right-hand columns, plasticity produces qualitative differences in the propagating wave: the leading element of the plastic NSH chain buckles and undergoes significant yielding, such that the rest of the chain does not buckle, while the all-elastic NSH elements experience a sequence of buckling events followed by a return to their undeformed configurations. Thus, the inclusion of plasticity in the DEM (or lack thereof) produces significant differences, both qualitative and quantitative, in the force transmitted through the chain.

\section{Conclusion}\label{sec:Conclusion}
	In this work, we have developed a DEM framework for simulation of 1D, nonlinear wave propagation in NLEM undergoing plastic deformation. By implementing elasto-plastic constitutive relations in the form of general dynamic constraints on the equations of motion, the framework enables rapid numerical simulations of rate- and history-dependent NLEM behavior for a wide variety of models. We have demonstrated this framework using four physically and mathematically distinct models, including one that was empirically derived from fine-scale FEA of a buckling lattice structure, and shown significant differences in the simulated dynamic behavior when plasticity is included or neglected.

	While the present work has been focused on 1D models that yield only under compressive loading, the generality of the method permits, in principle, extensions to more complex loading scenarios and higher dimensions. Future work should therefore address remaining issues toward this end, such as computational bottlenecks and the construction of constitutive relations for history-dependent deformation of complex structures in higher dimensions.	For example, since this framework generates DAEs of motion that generally require implicit time-stepping algorithms (such as the IRK methods chosen for the present work), computational costs can potentially be reduced by implementing automatic differentiation routines to more efficiently evaluate the Jacobian matrices required at each time step. Further gains in computational efficiency may be achieved by investigating time-stepping methods tailored for non-smooth constitutive relations, such as those used by the multi-body dynamics community to simulate collisions between rigid bodies \cite{stewart1996implicit}. Finally, while a brute-force curve-fitting approach was sufficient to generate a 1D empirical model for the NSH geometry studied herein, more advanced data driven techniques (\eg, machine learning or energetically equivalent surrogate models \cite{rossi2023surrogate}) will likely be required in higher dimensions.

\section{Acknowledgments}
	The authors acknowledge financial support from Kansas City National Security Campus (PDRD No. 705020).
\appendix

\section{Numerical Solution of the Differential-Algebraic Equations of Motion}\label{app:DAE}

	The DAEs of motion derived in this work, \Eqref{eq:eomChain}, may be written in the implicit, vectorial form
	\begin{equation}
		\vect{\Phi}(t,\vect{z},\dot{\vect{z}})=\vect{0},
	\end{equation} 
	where $\zz = (\vect{x},\vect{v},\vect{f},\vect{e},\vect{s},\vect{\upxi})$ is a block vector containing the $6N$ dependent variables (\ie, $\vect{x} = (x_1,x_2,\dots,x_N)$, \etc). DAEs are known to have high numerical stiffness and require time-stepping algorithms with excellent stability properties, such as implicit Runge-Kutta (IRK) methods and backward differentiation formulas \cite{HairerWannerBookII}, which are single- and multi-step methods, respectively. Since the elasto-plastic constitutive relations are generally non-smooth, we require a method that does not rely on continuity of the derivatives of the DAEs across time steps (\ie, we require a single-step method). Specifically, we select the {Radau-IIa} family of IRK methods \cite{Hairer1999}, which has been shown to perform well for DAEs modeling a wide variety of physical systems \cite{FabienBook}. While the inner workings of IRK methods are well-documented and generally outside the scope of the present study, the non-smooth nature of our specific DAEs necessitates a custom implementation, which deviates from those typically found in commercial and open-source packages \footnote{Typical implementations are optimized for large systems of smooth DAEs, such as those obtained via discretization of diffusive partial differential equations, and utilize higher-order polynomial extrapolation and infrequent Jacobian updates.}. In particular, for each time step, we compute the initial guess of the solution using linear extrapolation via the final stage derivative of the previous step, and solve the system of nonlinear algebraic equations using the Newton-Raphson method with the Jacobian updated at every iteration.
	
	For the simulation results presented in the main text, we use the 3-stage {Radau-IIa} method, which achieves 5th-order truncation error for smooth systems of ODEs and Index-1 DAEs. While this accuracy is not guaranteed for the non-smooth systems studied herein, we have ensured that time steps are sufficiently small to achieve convergence.

%

\end{document}